\documentclass[twocolumn,aps,pra,showpacs,superscriptaddress,floatfix]{revtex4}
\usepackage{graphicx}
\usepackage{times}
\usepackage{nicefrac}
\usepackage{amsmath}
\usepackage{amsfonts}
\usepackage{amssymb}
\usepackage{amsthm}
\usepackage{bm}

\begin{document}

\title{One-loop binding corrections to the electron $g$ factor}

\author{Krzysztof Pachucki}
\affiliation{Faculty of Physics, University of Warsaw, Pasteura 5, 02-093 Warsaw, Poland}

\author{Mariusz Puchalski} 
\affiliation{Faculty of Chemistry, Adam Mickiewicz University,
             Umultowska 89b, 61-614 Pozna{\'n}, Poland}

\begin{abstract}
  We calculate the one-loop electron self-energy correction of order $\alpha\,(Z\,\alpha)^5$
  to the bound electron $g$ factor. Our result is in agreement
  with the extrapolated numerical value and paves the way for the calculation of the analogous,
  but as yet unknown two-loop  correction.
\end{abstract}

\maketitle

\section{Introduction}

The $g$ factor of a bound electron  is the coupling 
constant of the spin to an external, homogeneous magnetic field.
In natural units $\hbar = c = \varepsilon_0 = 1$,
it is defined by the relation
\begin{equation}
\delta E = -\frac{e}{2\,m}\,
\langle \vec{\sigma} \vec{B} \rangle \,\frac{g}{2}\,,
\label{01}
\end{equation}
where $\delta E$ is the energy shift of the electron due to the
interaction with the magnetic field $\vec{B}$, 
$m$ is the mass of the electron, 
and $e$ is the electron charge ($e<0$).
It was found long ago~\cite{breit:28} that
in a relativistic (Dirac) theory,
the $g$ factor of a bound electron
differs from the value $g=2$ due to the so-called binding corrections.
For an $nS$ state, they are given by
\begin{align}
g &= \frac23 \left( 1+2\, \frac{E}{m}\right) 
\nonumber\\
&= 2-\frac{2}{3}\,\frac{(Z\,\alpha)^2}{n^2}+
\biggl(\frac{1}{2\,n}-\frac{2}{3}\biggr)\,\frac{(Z\,\alpha)^4}{n^3} +
\ldots \,,
\label{02}
\end{align}
where $E$ is the Dirac energy. In addition, there are many QED corrections,
and the dominant one comes from the so-called electron self-energy.
When expanded in powers of $Z\,\alpha$ the one-loop
electron self-energy correction reads (for the $nS$ state)
\begin{align}
g_{\rm SE} = \frac{\alpha}{\pi}\biggl[&
  1+\frac{(Z\,\alpha)^2}{6\,n^2} + \frac{(Z\,\alpha)^4}{n^3}
  \biggl(\frac{32}{9}\ln[(Z\,\alpha)^{-2}] + b_{40}(n)\!\biggr)\nonumber \\&
  +\frac{(Z\,\alpha)^5}{n^3}\,b_{50} + \frac{(Z\,\alpha)^6}{n^3}
  \Bigl( b_{62}\,\ln^2[(Z\,\alpha)^{-2}]\nonumber \\&
  + b_{61}(n)\,\ln[(Z\,\alpha)^{-2}] +b_{60}(n)\Bigr) +\ldots\biggr],\label{03}
\end{align}
where $b_{40}(1S) = -10.236\,524\,32$ \cite{pachucki:04:prl,pachucki:05:gfact} , $b_{50} = 23.6(5)$ \cite{yerokhin:17},
and higher order coefficients remains unknown.
What is approximately known, however, is the sum of $b_{50}$ and higher-orders terms
for individual nuclear charges from all-order numerical calculations
\cite{persson:97:g, blundell:97, beier:00:pra, yerokhin:17}.
The subject of this work is the one-loop electron self-energy correction
of the order of $\alpha\,(Z\,\alpha)^5$, namely the coefficient $b_{50}$. Although it has been
obtained by extrapolation of numerical results,
we aim to calculate it directly, in order to find out the best approach for the analogous
two-loop contribution, which currently is the main source of the uncertainty of
theoretical predictions.
Due to extremely accurate measurements in hydrogenlike carbon \cite{sturm:14},
the bound electron $g$ factor is  presently used for the most accurate determination
of the electron mass \cite{mohr:16:codata},
and in the future it can be used for determination of the fine structure constant \cite{yerokhin:16}
and for precision tests of the Standard Model.

\section{$\alpha\,(Z\,\alpha)^5$ correction to the Lamb shift}
\label{sec:1}
Before turning to the $g$ factor we present a simple derivation of the analogous correction
to the Lamb shift as proof of concept because the computational approach for the $g$ factor
will be very similar. The one--loop electron self-energy contribution to the Lamb shift is
\begin{equation}
  E_{\rm SE} = e^2\,\int\frac{d^4k}{(2\,\pi)^4\,i}\,\frac{1}{k^2}\,
  \langle\bar\psi|\gamma^\mu\,\frac{1}{\not\!p+\not\!k-\gamma^0\,V-m}\,\gamma_\mu|\psi\rangle,
  \label{04}
\end{equation}
where $V = -Z\,\alpha/r$.
The $(Z\,\alpha)^5$ contribution is obtained from the hard two-Coulomb exchange
\begin{eqnarray}
  E_{\rm SE}^{(5)} &=&e^2\,\phi^2(0)\,(Z\,\alpha)^2\,\int\frac{d^3q}{(2\,\pi)^3}\,\frac{f(\vec q^{\,2})}{\vec q^{\,4}},
  \label{05}\\
  f(\vec q^{\,2}) &=& \int\frac{d^4k}{i\,\pi^2}\frac{1}{k^2}
  {\rm Tr}\biggl[\bigl(T_1+2\,T_2+T_3\bigr)\,\biggl(\frac{\gamma^0+I}{4}\biggr)\biggr], \label{06}
\end{eqnarray}  
where
\begin{align}
  T_1 =& \gamma^\mu\,\frac{1}{\not\!t+\not\!k-m}\,\gamma^0\,
  \frac{1}{\not\!t+\not\!k+\not\!q-m}\,\gamma^0\,
  \frac{1}{\not\!t+\not\!k-m}\,\gamma_\mu,  \nonumber \\
  T_2 =& \gamma^0\,\frac{1}{\not\!t+\not\!q-m}\,\gamma^\mu\,
  \frac{1}{\not\!t+\not\!k+\not\!q-m}\,\gamma^0\,
  \frac{1}{\not\!t+\not\!k-m}\,\gamma_\mu,  \nonumber \\
  T_3 =& \gamma^0\,\frac{1}{\not\!t+\not\!q-m}\,\gamma^\mu\,
  \frac{1}{\not\!t+\not\!k+\not\!q-m}\,\gamma_\mu\,
  \frac{1}{\not\!t+\not\!q-m}\,\gamma^0,  \nonumber\\ \label{07}
\end{align}
and where $t = (m,0,0,0)$, $t\,q = 0$, $q^2 = -\vec q^{\,2}$.
Equation (\ref{05}) as it stands is divergent at small $\vec q^{\,2}$. One subtracts
leading terms in small $\vec q^{\,2}$, which correspond to lower order contributions to the Lamb shift,
so $f(\vec q^{\,2}) \sim \vec q^{\,2}$, and 
\begin{equation}
f(\vec q^{\,2}) = \vec q^{\,2}\,\int d(p^2)\,\frac{1}{p^2\,(\vec q^{\,2}+p^2)}\,f^A(p^2)\label{08}
\end{equation}
function $f$ can be expressed in terms of its imaginary part $f^A$ on a cut $\vec q^{\,2}<0$ 
\begin{equation}
f^A(p^2) = \frac{f(-p^2+i\,\epsilon) - f(-p^2-i\,\epsilon)}{2\,\pi\,i}.\label{09}
  \end{equation}
The correction to energy in terms of $f^A$ becomes
\begin{equation}
  E_{\rm SE}^{(5)} =e^2\,\phi^2(0)\,(Z\,\alpha)^2\,\int\frac{d\,p}{2\,\pi}\,\frac{f^A(p^{2})}{p^{2}}.
  \label{10}
\end{equation}
The imaginary part $f^A$ is much easier to evaluate because it does not involve any infrared or ultraviolet
divergences in $k$ and has much simpler analytic form than the $f$ itself.
The calculations go as follows. Traces are performed with {\sl FeynCalc} package \cite{feyncalc}.
The resulting expression is a linear combination of fractions with the numerator containing powers
of $k^2, q^2, k\,t$, and $k\,q$,
while $q\,t$ vanishes. Any $k$ in the numerator can be reduced with the denominator with the help of
\begin{eqnarray}
  k\,q &=& \frac{1}{2}\,\bigl[(k+q+t)^2 - (k+t)^2 - q^2\bigr]\,,\\
  k\,t &=& \frac{1}{2}\,\bigl[(k+t)^2 - k^2 - q^2\bigr]\,.\nonumber
\end{eqnarray}
The resulting expression is a linear combination of
\begin{equation}
  \frac{1}{i\,\pi^2} \int d^4\,k\frac{1}{[k^2]^n\,[(k+t)^2-1]^m\,[(k+t+q)^2-1]^l}
\end{equation}
with integer $n,m,l \geq 0$.
Next, the powers $n,m,l$ are reduced to 1 or 0 using integration by parts identities
\begin{equation}
  \int d^4\,k\,\frac{\partial}{\partial k^\mu} \frac{p^\mu}{[k^2]^n\,[(k+t)^2-1]^m\,[(k+t+q)^2-1]^l} = 0\\
\end{equation}
with $p = k,q,t$. The resulting expression contains the integral
\begin{equation}
  J =   \frac{1}{i\,\pi^2} \int d^4\,k\frac{1}{k^2\,[(k+t)^2-1]\,[(k+t+q)^2-1]}
\end{equation}
and simpler integrals without any of these denominators.
Analytic expressions for all such integrals can be taken from \cite{thooft}, but it is much easier
to calculate the imaginary part using Feynman parameters. For example, the imaginary part of
the $J$-integral is
\begin{equation}
  J^A(p^2) = \frac{1}{p}\,\biggl[\arctan(p) - \Theta(p-2)\,\arccos\biggl(\frac{2}{p}\biggr)\biggr]\,.
\end{equation}
Using $J^A$ and simpler formula for other integrals the result for $f^A$ is
\begin{eqnarray}
  f^A(p^2) &=& \frac{7}{3} - \frac{16}{p^2} - \frac{1}{1 + p^2} +
  \biggl(\frac{16}{p^3} + \frac{4}{p} - p\biggr)\,\arctan(p)\nonumber \\ &&
  +4\,\biggl(1 + \frac{1}{p^2} - \frac{12}{p^4} \biggr)\,\frac{\Theta(p-2)}{\sqrt{1-4/p^2}}\nonumber \\ &&
  - \biggl(\frac{16}{p^3} + \frac{4}{p} - p\biggr)\,\Theta(p-2)\,\arccos\biggl(\frac{2}{p}\biggr).
  \label{11} 
\end{eqnarray}
The one dimensional integration in Eq. (\ref{10}) leads to
\begin{equation}
\int\frac{d\,p}{2\,\pi}\,\frac{f^A(p^{2})}{p^{2}} = \frac{139}{128}- \frac{\ln 2}{2} \equiv C. \label{12}
 \end{equation}
Finally, the result for the $\alpha\,(Z\,\alpha)^5$ electron self-energy contribution to the Lamb shift
\begin{equation}
  E_{\rm SE}^{(5)} = m\,\frac{\alpha\,(Z\,\alpha)^5}{n^3}\,4\,C, \label{13}
\end{equation}
is in agreement with the well-known value \cite{mohr:16:codata, eides}.
The same integration technique is used in the next paragraph for the evaluation
of the analogous correction to the $g$ factor.

\section{$\alpha\,(Z\,\alpha)^5$ correction to the $g$ factor}

The one-loop correction to the g factor is similar to Eq. (\ref{04})
\begin{equation}
  \delta E = e^2\!\int\!\frac{d^4k}{(2\,\pi)^4 i}\frac{1}{k^2}
  \langle\bar\psi|\gamma^\mu\frac{1}{\not\!p+\not\!k-e\,\not\!\!A-\gamma^0 V-m}\gamma_\mu|\psi\rangle
  \label{14}
\end{equation}
where $\psi$ is the electron wave function which includes perturbation due to
external magnetic field $A$, and
$p^0$ includes the corresponding energy shift
\begin{equation}
p_0 = E+\langle\bar\psi|e\not\!\!A|\psi\rangle. \label{15}
  \end{equation}
The $(Z\,\alpha)^5$ contribution is given in analogy to the Lamb shift,
by the hard two-Coulomb exchange
\begin{widetext}
\begin{align}
  \delta E^{(5)} = e^2\,\int\frac{d^4k}{(2\,\pi)^4\,i}\,\frac{1}{k^2}\,\biggl\langle
  \bar\psi\biggl|&\gamma^\mu\,\frac{1}{\not\!p+\not\!k-e\,\not\!\!A-m}\,\gamma^0\,V\,
  \frac{1}{\not\!p+\not\!k-e\,\not\!\!A-m}\,\gamma^0\,V\,\frac{1}{\not\!p+\not\!k-e\,\not\!\!A-m}\,
  \gamma_\mu
  \nonumber \\ &
+2\,\gamma^0\,V\,\frac{1}{\not\!p-e\,\not\!\!A-m}\,\gamma^\mu\,
  \frac{1}{\not\!p+\not\!k-e\,\not\!\!A-m}\,\gamma^0\,V\,\frac{1}{\not\!p+\not\!k-e\,\not\!\!A-m}\,
  \gamma_\mu\nonumber \\ &
+\gamma^0\,V\,\frac{1}{\not\!p-e\,\not\!\!A-m}\,\gamma^\mu\,
  \frac{1}{\not\!p+\not\!k-e\,\not\!\!A-m}\,\gamma_\mu\,\frac{1}{\not\!p+\not\!k-e\,\not\!\!A-m}\,\gamma^0\,V
  \biggr|\psi\biggr\rangle, \label{16}
\end{align}
\end{widetext}
and by the expansion in $A$ and in the momentum carried by $A$.
The expansion of $\psi$ in $A$ is not very trivial.
Since only the low momenta of the wave function $\psi$ contribute to $(Z\,\alpha)^5$
we apply the Foldy-Wouthyusen transformation in the presence of the magnetic field
\begin{equation}
S = -\frac{\rm i}{2\,m}\,\vec\gamma\cdot\vec\pi, \label{17}
\end{equation}
and the wave function can be represented as
\begin{equation}
  |\psi\rangle =   e^{-{\rm i}\,S}\,
  \left|\begin{array}{c} \phi\\ 0 \end{array}\right\rangle
  = \biggl(I-\frac{1}{2\,m}\,\vec\gamma\,\vec\pi +\frac{e}{8\,m^2}\,\vec\sigma\vec B\biggr)
  \left|\begin{array}{c} \phi\\ 0 \end{array}\right\rangle, \label{18}
  \end{equation}
where $\phi$ is the spinor wave function which corresponds to the transformed Hamiltonian
\begin{align}
H' =& e^{{\rm i}\,S}\,(H-i\,\partial_t)\,e^{-{\rm i}\,S} \nonumber \\
   =& \frac{p^2}{2\,m} -\frac{Z\,\alpha}{r} -\frac{e}{2\,m}\,\vec\sigma\vec B
  \biggl(1-\frac{p^2}{2\,m^2}+\frac{Z\alpha}{6\,m\,r}\!\biggr).\label{19}
\end{align}
We are now ready to perform an expansion in $\not\!\!A$ of Eq. (\ref{16}),
and split $\delta E^{(5)}$ in four parts
\begin{equation}
  \delta E^{(5)} = E_1 + E_2 + E_3+E_4\,. \label{20}
\end{equation}

$E_1$ comes from the last term in Eq. (\ref{18})
\begin{equation}
  E_1 = \frac{e}{4\,m^2}\,\,\langle\vec\sigma\cdot\vec B\rangle\; E^{(5)} =
  -\frac{e}{2\,m}\,\langle\vec\sigma\cdot\vec B\rangle\,\frac{g_1}{2}, \label{21}
\end{equation}
where
\begin{equation}
  g_1 = -\frac{E^{(5)}}{m} = -\frac{\alpha\,(Z\,\alpha)^5}{n^3}\,4\,C\,.\label{22}
\end{equation}

$E_2$ comes from perturbation of $\phi$ due to the last term in the transformed Hamiltonian (\ref{19})
\begin{equation}
  E_2 = \frac{e}{m}\,\langle\vec\sigma\cdot\vec B\rangle\,C\,\alpha\,(Z\,\alpha)^5
  \biggl\langle\frac{5}{6\,r}\frac{1}{(E-H)'}\,4\,\pi\,\delta^{(3)}(r)\biggr\rangle, \label{23}
\end{equation}
where $p^2/2$ is replaced by $1/r$. Since
\begin{equation}
  \frac{1}{(E-H)'}\, \frac{1}{r}\,\phi = -\frac{\partial}{\partial\alpha}\,\phi, \label{24}
\end{equation}
the above matrix element is
\begin{equation}
\biggl\langle\frac{1}{r}\frac{1}{(E-H)'}\,4\,\pi\,\delta^{(3)}(r)\biggr\rangle = -\frac{6}{n^3}\,, \label{25} 
\end{equation}
and $g_2$ becomes
\begin{equation}
g_2 = \frac{\alpha\,(Z\,\alpha)^5}{n^3}\,20\,C\,.\label{26}
  \end{equation}

$E_3$ comes from expansion of Eq. (\ref{16}) in $p_0-m=-e\,\langle\vec\sigma\vec B\rangle/(2\,m)$,
\begin{equation}
  E_3 = -\frac{e}{2\,m}\,\langle\vec\sigma\cdot\vec B\rangle\,
  e^2\,\phi^2(0)\,(Z\,\alpha)^2\,C'\,, \label{27}
\end{equation}
where
\begin{eqnarray}
  C'&=&\frac{\partial}{\partial E}\biggr|_{E=1}
  \int\frac{d^3q}{(2\,\pi)^3}\,\frac{1}{\vec q^{\,4}}
  \int\frac{d^4k}{i\,\pi^2}\,\frac{1}{k^2} \nonumber\\&&
  \times{\rm Tr}\biggl[\bigl(T_1+2\,T_2+T_3\bigr)\,\biggl(\frac{\gamma^0+I}{4}\biggr)\biggr] \label{28}
  \\&=&  -\frac{659}{256} + \ln(2)\,,    \nonumber   
\end{eqnarray}  
and where $T_i$ are defined in Eq. (\ref{07}) with $t=(E,0,0,0)$.
The corresponding correction to the $g$ factor is
\begin{equation}
g_3 = \frac{\alpha\,(Z\,\alpha)^5}{n^3}\,8\,C'\,.\label{29}
\end{equation}

The last term  $E_4$ comes from the expansion of $\delta E^{(5)}$ in $\vec\gamma\cdot\vec A$.
A typical contribution is of the form
\begin{widetext}
\begin{align}
  E_4 &= e^2\,\int\frac{d^4k}{i\,\pi^2}\,\frac{1}{k^2}\,
  \int\frac{d^3p}{(2\,\pi)^3}\,\frac{Z\,\alpha}{(-\vec p-\vec q/2)^2}
  \frac{Z\,\alpha}{(\vec p-\vec q/2)^2}\,\phi^2(0)\,e\,i\,\epsilon^{ijk}\,\sigma^k
  \nonumber \\&
  {\rm Tr}\biggl[
  \gamma^\mu\,\frac{1}{\not\!t+\not\!k-m}\,\gamma^0\,
  \frac{1}{\not\!t+\not\!p+\not\!q/2+\not\!k-m}\,\not\!\!A(q)\,
  \frac{1}{\not\!t+\not\!p-\not\!q/2+\not\!k-m}
  \,\gamma^0\,\frac{1}{\not\!t+\not\!k-m}
  \,\gamma_\mu\,\frac{(\gamma^0+I)}{16}\,[\gamma^i\,,\,\gamma^j]\biggr]+\ldots\label{30}
\end{align}
\end{widetext}
where by dots we denote all other diagrams.
In addition, we perform an expansion in the momentum $\vec q$ transferred by $A$ and obtain
\begin{eqnarray}
E_4 &=& e^2\,(Z\,\alpha)^2\,\phi^2(0)\,C''\,(A^i\,q^j-A^j\,q^i)\,e\,i\,\epsilon^{ijk}\,\sigma^k\nonumber \\
&=& -2\,e^2\,(Z\,\alpha)^2\,\phi^2(0)\,C''\,e\,\vec\sigma\,\vec B, \label{31}
\end{eqnarray}
where
\begin{equation}
  C'' = \frac{281}{1024} + \frac{\ln(2)}{12}\,. \label{32}
\end{equation}
The corresponding correction to the $g$ factor is
\begin{equation}
g_4 = \frac{\alpha\,(Z\,\alpha)^5}{n^3}\,32\,C''\,. \label{33}
\end{equation}
The total $\alpha\,(Z\,\alpha)^5$ contribution to the bound electron $g$ factor is
the sum of individual corrections, namely
\begin{eqnarray}
  g^{(5)} &=& g_1 + g_2 + g_3 + g_4 \nonumber\\
  &=& \frac{\alpha\,(Z\,\alpha)^5}{n^3}\,\bigl(16\,C + 8\,C' + 32\, C''\bigr) \label{34} \\
  &=&  \frac{\alpha\,(Z\,\alpha)^5}{n^3}\,\biggl(\frac{89}{16} + \frac{8\,\ln(2)}{3}\biggr). \nonumber
\end{eqnarray}
The numerical value for the coefficient multiplied by $\pi$ is $b_{50} = 23.282\,005$,
in agreement with  Yerokhin's very recent result of $23.6(5)$ \cite{yerokhin:17}.
However, what is not in agreement is the difference for $b_{50}(2S) - b_{50}(1S)$,
which according to our calculations vanishes, but Yerokhin {\em et al.} \cite{yerokhin:17}
give $0.12(5)$. All the assumptions in performing the fit in Ref. \cite{yerokhin:17}
were correct, so this small discrepancy needs further investigation.

\section{Summary}
We have calculated the one-loop electron self-energy contribution of order $\alpha\,(Z\,\alpha)^5$
to the bound electron $g$ factor, and found that it is state independent. The principal result, however,
is a presentation of the computational approach, which can be extended to
the yet unknown two-loop correction. This correction is presently the main source of theoretical uncertainty.
The extension of the direct one-loop numerical calculation  to the two-loop case is presently
out of reach. In contrast, the analytic approach with an expansion in $Z\,\alpha$
is technically as difficult as the two-loop self-energy correction to the Lamb shift,
which has been known for some time \cite{eides}.

\begin{acknowledgments}
This work was supported by National Science Center (Poland) Grant
No. 2012/04/A/ST2/00105.
\end{acknowledgments}


\end{document}